\documentclass[aip,reprint]{revtex4-2}

\usepackage{algorithm}
\usepackage{algpseudocode}
\usepackage{lipsum}
\usepackage{amsmath}
\usepackage{amssymb}
\usepackage{graphicx}
\usepackage{gensymb}
\usepackage{tabularx}
\usepackage{float}
\floatstyle{plain}
\restylefloat{table}  
\usepackage{xcolor}
\usepackage{mathtools}
\usepackage{hyperref} 
\usepackage[capitalise]{cleveref}
\DeclareUnicodeCharacter{2009}{ }
\DeclareUnicodeCharacter{2192}{$\rightarrow$}
\DeclareUnicodeCharacter{2032}{'}
\DeclareUnicodeCharacter{03B1}{$\alpha$}
\DeclareUnicodeCharacter{03B3}{$\gamma$}
\restylefloat{table}
\graphicspath{ {images/} }

\begin{document}
	\title{Automated simulation-based design via multi-fidelity active learning and optimisation for laser direct drive implosions}
	\author{A. J. Crilly}\email{ac116@ic.ac.uk}
	\affiliation{Centre for Inertial Fusion Studies, The Blackett Laboratory, Imperial College, London SW7 2AZ, United Kingdom}
    \affiliation{I-X Centre for AI In Science, Imperial College London, White City Campus, 84 Wood Lane, London W12 0BZ, United Kingdom}
	\author{P. W. Moloney}
	\affiliation{Centre for Inertial Fusion Studies, The Blackett Laboratory, Imperial College, London SW7 2AZ, United Kingdom}
    \author{D. Shi}
	\affiliation{Centre for Inertial Fusion Studies, The Blackett Laboratory, Imperial College, London SW7 2AZ, United Kingdom}
    \author{E. A. Ferdinandi}
	\affiliation{Centre for Inertial Fusion Studies, The Blackett Laboratory, Imperial College, London SW7 2AZ, United Kingdom}

    \begin{abstract}
    The design of inertial fusion experiments is a complex task as driver energy must be delivered in a precise manner to a structured target to achieve a fast, but hydrodynamically stable, implosion. Radiation-hydrodynamics simulation codes are an essential tool in this design process. However, multi-dimensional simulations that capture hydrodynamic instabilities are more computationally expensive than optimistic, 1D, spherically symmetric simulations which are often the primary design tool. In this work, we develop a machine learning framework that aims to effectively use information from a large number of 1D simulations to inform design in the presence of hydrodynamic instabilities. We use an ensemble of neural network surrogate models trained on both 1D and 2D data to capture the space of good designs, i.e. those that are robust to hydrodynamic instabilities. We use this surrogate to perform Bayesian optimisation to find optimal designs for a 25 kJ laser driver. We perform hydrodynamic scaling on these designs to confirm the achievement of high gain for a 2 MJ laser driver, using 2D simulations including alpha heating effects.
    \end{abstract}

    \maketitle
    
    \section{Introduction}

    Inertial confinement fusion (ICF) seeks to compress and heat deuterium–tritium (DT) fuel to thermonuclear conditions by coupling driver energy into a spherically imploding shell \cite{Atzeni2004,lindl1995}. In laser direct drive\cite{craxton2015direct} (LDD), multiple laser beams directly illuminate a capsule, generating an ablative rocket that accelerates the shell inward. Achieving ignition\cite{Abu2022,Zylstra2022ign,Kritcher2022ign} and appreciable burn propagation requires a careful orchestration of target dimensions and laser pulse history to control shock timing, adiabat, and implosion velocity, while mitigating hydrodynamic instabilities \cite{Betti2016,Hurricane2014}. These competing considerations create a high-dimensional, strongly coupled design space in which small changes to one parameter often necessitate compensating changes in other design parameters. 
    
    Radiation–hydrodynamics (rad-hydro) simulations are indispensable for navigating this space, but their computational cost grows steeply with dimensionality. One-dimensional (1D) simulations enable large surveys and provide an upper bound on performance through perfect spherical symmetry, whereas multi-dimensional (2D/3D) simulations are needed to capture Rayleigh–Taylor instability growth that degrades fusion yield\cite{Haines2020,Clark2019,McGlinchey2018}. However, the computational cost of 2D calculations can exceed that of 1D by two orders of magnitude, and 3D simulations are more expensive still.
    
    These characteristics motivate multi-fidelity (1D, 2D or 3D) and active-learning machine learning strategies that extract as much information as possible from inexpensive low-dimensional simulations while judiciously deploying costly high-dimensional simulations. Surrogate models can learn mappings from design variables to outcomes and provide predictive uncertainties for Bayesian optimisation \cite{Snoek2012}. Multi-fidelity extensions exploit correlations across fidelities to accelerate convergence \cite{KennedyOHagan2000,Perdikaris2017}. However, standard multi-fidelity formulations often assume modest computational cost ratios between the fidelities and noise models that are uniform across the fidelities. Thus for multi-fidelity ICF design, there is a need for methods that (i) respect extreme cost disparities, (ii) remain robust to non-uniform, fidelity-dependent noise, and (iii) integrate seamlessly with pre-existing rad-hydro codes; which require High Performance Computing (HPC) clusters.
    
    In this work we develop an automated, simulation-based design framework that couples multi-fidelity surrogate modelling with active learning and Bayesian optimisation for LDD implosions. Our contributions are:
    \begin{enumerate}
    \item A 1D and 2D radiation-hydrodynamics simulation pipeline allowing straightforward coupling to machine learning (ML) models. For 2D simulations we include fixed (beam-mode) and randomised (ablator surface structure)  perturbation sources to reflect shot-to-shot variability.
    \item Multi-fidelity surrogate models that transfer information from large 1D ensembles to data-scarce 2D. These include a neural network ensemble with transfer learning and calibrated predictive uncertainties, alongside Gaussian process baselines.
    \item An active-learning scheme based on probabilistic threshold sampling that concentrates expensive simulations in promising regions of the design space.
    \item Bayesian optimisation at fixed fidelity using multi-fidelity-trained surrogates to find optimal designs in both 1D and 2D. We define optimality via a physics-informed objective that blends an ignition criterion\cite{Betti2015,Christopherson2020}, $\chi_{\mathrm{no}\ \alpha}$, with post-ignition burn propagation via areal density\cite{Fraley1974} $\rho R$, enabling sub-scale optimisation that targets ignition and high gain at a larger hydrodynamic scale.
    \end{enumerate}
    
    We demonstrate the framework on an OMEGA-relevant LDD problem with the OMEGA\cite{Boehly1997} 60-beam geometry, 25kJ driver and an eight-parameter design space (laser picket/foot/rise and layered CH/DT target). We first amass a large 1D database to train surrogates and narrow down the search domain. We then conduct a targeted 2D campaign informed by active learning. We find that transferring structure learned in 1D dramatically reduces the number of 2D simulations required to locate robust designs. Importantly, we show that the 2D-optimised design exhibits substantially improved hydrodynamic stability compared to the 1D-optimised counterpart at the same energy scale. Finally, we assess performance under hydrodynamic scaling\cite{Nora2014} to 2MJ (National Ignition Facility\cite{Miller2004}-scale) with both burn-off (no alpha heating) and burn-on (with alpha heating) simulations to gauge ignition margin and post-ignition burn.
    
    The remainder of the paper is organised as follows. Section II presents the methodology, defining the design space, objective, and simulation models and detailing the surrogate, active-learning, and optimisation components. Section III presents the optimisation studies in 1D and 2D. Section IV reports the performance of the optimal designs at 25kJ and at 2MJ with burn-off/on analyses. Future work and conclusions are then discussed.
    
    \section{Methodology}

    \subsection{Design}

    Design of laser direct drive ICF implosions involves the prescription of a laser pulse (incident laser power as a function of time) and a target (spherical layered target dimensions)\cite{craxton2015direct}. Each of the design parameters (both individually and in combination) are strongly linked to behaviours of the implosion. By tuning of the parameters, one can alter the implosion design (implosion velocity, shell mass, etc.) to achieve the desired conditions at stagnation. However, this is no simple task due to the complex non-linear relationships created by the many physicial processes inherent to ICF. Human designers must consider the qualitative effect of each design decision and use simulations as a tool to quantify these effects. In this work we look to develop a framework that can learn these relationships and thus autonomously move towards an optimal design.

    To proceed we must parameterise the laser pulse and target design. The full LDD design space is very high dimensional and therefore we look to reduce the problem down to $< 10$ parameters. Firstly, we consider LDD implosions with a 60-beam 25 kJ laser driver. This is relevant to the OMEGA laser facility\cite{Boehly1997}, allowing comparisons between the designs presented in this work with the large OMEGA experimental and simulation database. For the pulse design, we consider a single picket, a foot and power law rise to peak power ($P_{\mathrm{peak}}$ = 25 TW). For the power law rise we use a simplified form of the functional prescription in Gopalaswamy \textit{et al.}\cite{gopalaswamy2025automated}:
    \begin{subequations}
    \begin{align}
        P(t) &= P_{\mathrm{picket}}(t) + P_{\mathrm{main}}(t) \ , \\
        P_{\mathrm{picket}}(t) &= P_{\mathrm{picket}} \exp\left[-\frac{1}{2}\left(\frac{t-3\sigma_t}{\sigma_t}\right)^2\right] \, \\
        P_{\mathrm{main}}(t) &= 
        \begin{cases}
        P_{\mathrm{foot}}\frac{\mathrm{erf}(x)+1}{2} & \text{for} \ \tau < 0 \\
        P_{\mathrm{foot}}\left[1-\xi \tau\right]^{-\beta} & \text{for} \ 0 < \tau < 1 \\
        P_{\mathrm{peak}}  & \text{for} \ \tau > 1 \ \& \ t < t_{\mathrm{end}}
        \end{cases} \ , \\
        \tau &= \frac{t-(t_{\mathrm{foot}}+t_{\mathrm{delay}}+3\sigma_t)}{t_{\mathrm{rise}}}\ , \\
        x &= \frac{t-(t_{\mathrm{delay}}+3\sigma_t)}{\sqrt{2} \sigma_t} \ , \\
        \xi &= 1 - \left(\frac{P_{\mathrm{peak}}}{P_{\mathrm{foot}}}\right)^{-\frac{1}{\beta}} \ , 
    \end{align}
    \end{subequations}
    Where $\sigma_t = 25$ ps and $\beta$ = 9/5. For the target, we use an undoped CH ablator layer, a DT ice layer and a central DT gas region; parameterised as two thicknesses and an outer radius. This laser and target design has a total of 8 free parameters. We must also pick acceptable ranges for these parameters, and for the target parameters we must pick discrete increments such that the initial target fits precisely on an Eulerian grid. \cref{tab:design} details the design space, note that all pulse duration parameters were capped at 1 ns to prevent excessively long pulse durations.

    \begin{table}[h]
        \centering
        \begin{tabular}{c|c|c|c}
        Variable (units) & Min & Max & Step size \\
        \hline
            $P_{\mathrm{picket}}$ (TW) & 0 & 25 & - \\
            $P_{\mathrm{foot}}$ (TW) & 0 & 12.5 & - \\
            $t_{\mathrm{delay}}$ (ps) & 50 & 1000 & - \\
            $t_{\mathrm{foot}}$ (ps) & 50 & 1000 & - \\
            $t_{\mathrm{rise}}$ (ps) & 50 & 1000 & - \\
            $R_{\mathrm{outer}}$ (um) & 400 & 650 & 1 \\
            $\Delta_{\mathrm{ablator}}$ (um) & 2 & 20 & 0.5 \\
            $\Delta_{\mathrm{ice}}$ (um) & 10 & 100 & 0.5 \\
        \end{tabular}
        \caption{Design space used in this work (8 total parameters, 5 laser, 3 target). See equations 1a-f for laser pulse parameterisation.}
        \label{tab:design}
    \end{table}

    With a design space chosen, one must define an objective function. An objective function is used to quantify progress towards some optimal state, where the objective function is maximised. Ultimately, the goal of ICF is to optimise fusion yield, however for sub-scale implosions, which cannot ignite, a different choice of objective is often needed. In the literature, it is common for sub-ignition scale implosions to optimise an ignition metric which can be hydrodynamically scaled\cite{Nora2014} to higher laser energies, such as $\chi_{no \ \alpha}$ \cite{Betti2015,Christopherson2020}. Once an ICF target has ignited, then burn propagation into the fuel mass begins. For burn propagation, the metric of success is the areal density ($\rho R$). This can be seen through Fraley's formula\cite{Fraley1974} for the fraction of DT fuel, $\Phi$, which undergoes fusion:
    \begin{equation}\label{eqn:Fraley}
        \Phi = \frac{\rho R}{6.3 \ \mathrm{g/cm}^2 + \rho R}
    \end{equation}
    We can therefore aim to design implosions at a lower energy scale which would ignite and achieve high gain at a higher energy scale. The ICF optimisation problem can be broken up into sub-problems to tune aspects of the implosion to achieve the overarching goal \cite{gopalaswamy2025automated}. However, we will focus on the global optimisation problem (as in Ref. \cite{Hatfield2019}) and therefore need only provide a single objective function. Guided by the need for ignition and burn propagation at the higher energy scale, we devised an objective which aims to smoothly combine criteria for ignition and burn propagation into a single scalar function:
    \begin{subequations}
    \begin{align}\label{eqn:objective}
        Y &= \min (\chi_{S, no \ \alpha} , 1)  \\ &+ \frac{1}{\hat{\rho R}} \max( \rho R - \rho R_{\chi = 1}, 0) \ , \nonumber \\
        \chi_{S, no \ \alpha} &=   S \left(\frac{\rho R}{1 \mathrm{g/cm}^2}\right)^{0.61} \left(0.12\frac{Y_{DT}}{10^{16}}\frac{1 \ \mathrm{mg}}{M_{\mathrm{stag}}}\right)^{0.34}\ , \\
        \rho R_{\chi = 1} &= \left(0.12\frac{Y_{DT}}{10^{16}}\frac{1 \ \mathrm{mg}}{M_{\mathrm{stag}}}\right)^{-1.64}\mathrm{g/cm}^2 \ , \\
        \hat{\rho R} &= 0.1 \ \mathrm{g/cm}^2 \ , \\
        S &= \left(\frac{E_{\mathrm{L,scaled up}}}{E_{\mathrm{L}}}\right)^{\frac{1}{3}}
    \end{align}
    \end{subequations}
    Where $S$ is the hydrodynamic scale factor, $Y_{DT}$ is the DT yield, and $M_{\mathrm{stag}}$ is the stagnated mass. The objective function aims to achieve ignition at scale $S$ (occurring when\cite{Christopherson2020} $\chi_{S, no \ \alpha} \sim 1$) before switching focus to increasing areal density above that required for ignition. In this work, we will consider scaling laser driver energies from 25 kJ (OMEGA scale) to 2 MJ (NIF scale). This leads to a scale factor $S \sim 4.3$.
    
    To calculate the quantities required for $\chi_{no \ \alpha}$ and $\rho R_{\chi = 1}$, we define the following concrete multi-dimensional definitions for implementation within the radiation-hydrodynamics simulations (in any dimensionality):
    \begin{subequations}
    \begin{align}
        \rho R &= \frac{1}{Y_{DT}} \int dt  \frac{d Y_{DT}}{dt} \int  \rho(r,\Omega,t) d\Omega dr \ , \\
        M_{\mathrm{stag}} &= \frac{1}{Y_{DT}} \int dt  \frac{d Y_{DT}}{dt} \int_{V_{\langle P \rangle e^{-2}}}  \rho(\vec{r},t) d^3r \ , \\
        \langle P \rangle &= \frac{1}{Y_{DT}} \int dt \int  P(\vec{r},t) \frac{d^2 Y_{DT}}{dt dV} d^3r
    \end{align}
    \end{subequations}
    Where $V_{\langle P \rangle e^{-2}}$ is the volume of fuel with pressures exceeding 2 e-foldings of the burn-averaged pressure, $\langle P \rangle$.

    \subsection{Radiation-hydrodynamics simulation}

    The design code used in this work was Chimera, a fixed-grid Eulerian radiation magnetohydrodynamics code, details of ICF implosion modelling with Chimera can be found in the literature\cite{Chittenden2016,Walsh2017,McGlinchey2018,Tong2019,Crilly2022,SpK2022}. To model the laser drive, Chimera was coupled to SOLAS, a 3D laser raytracing model, the details of which will be given in a future publication. Finally, a number of additional code developments were employed to speed up simulation time, namely implicit methods for both flux-limited thermal conduction and multi-group flux-limited radiation diffusion, physics for which Chimera has traditionally used explicit methods\cite{Meyer2014,SpK2022}. The atomic physics model SpK\cite{SpK2022} was used to generate multi-group opacity tables. To facilitate automation, control has been shifted to external input decks and standard decks developed for laser direct drive applications. 

    The purpose of 2D simulations is to capture hydrodynamic instability growth which degrades performance over the perfect spherical symmetry of 1D. The laser beam geometry introduces a seed for this instability growth in the form of low mode intensity variations on the capsule surface, known as beam mode\cite{Skupsky1983,Gopalaswamy2021}. Laser speckle operates at much shorter lengthscales than beam mode, introducing high mode sources of instability\cite{Skupsky1999,Goncharov2000,Patel2023}. Capsule defects, like surface roughness and voids, are also key seeds for instability growth for both low and high mode \cite{Shiau1974}. To capture some of these effects in our 2D simulations we (a) used the OMEGA beam port geometry to introduce beam mode and (b) introduced density perturbations on the outer 2.5um of the CH ablator. These density perturbations could have been kept constant across different simulations; however, we opted to reseed the random perturbations to introduce noise to the simulated outputs. This gives better correspondence to the reality of shot-to-shot variation in experiment and ensures our methods are robust to such noise. The level of perturbation introduced led to $\sim$1\% RMS areal density variations for the initial conditions.

    For multi-dimensional Eulerian simulations of spherical implosions it is beneficial to run in spherical geometry for as long as possible to prevent non-spherical grid-based artifacts. However, multi-dimensional regular spherical grids have small cell sizes towards r = 0, requiring small numerical time steps for stability. A restart to other geometries can be used to eliminate this restrictive time step requirement. For this work, an automatic restarting scheme was devised whereby initially 2D spherical ($r-\theta$) geometry was run up until the first shock approached r = 0. At this time, the variables for restart are dumped to disk and a restart is initialised onto a 2D cylindrical ($r-z$) grid. The outer extent of the cylindrical grid was found by identifying the outermost radius at a few percent of critical density. 

    The above code optimisations allowed run times of approximately 1 and 200 core hours for 1D and 2Ds respectively. For OMEGA scale implosions, a radial resolution of 500nm was used and 256 polar cells in 2D. The simulations were run on 32 and 128 CPU cores for 1D and 2D respectively.

    \subsection{Surrogate modelling}

    Surrogate models are a key component of machine learning and optimisation methods. These models use sample data to learn an approximate functional relationship between the inputs ($X$) and the outputs ($Y$). Once trained on this data, the surrogate model is much cheaper to sample from than creating new sample data, i.e. cheaper than running a rad-hydro simulation. Additionally, surrogate models which provide an uncertainty estimate in their prediction are particular useful, allowing Bayesian optimisation \cite{Snoek2012}.

    \subsubsection{Gaussian Processes}

    Gaussian processes (GPs) are a common surrogate model for optimisation purposes. There have been a number of works using GPs as surrogate models in ICF modelling and optimisation \cite{gopalaswamy2025automated,Wang2024}. 

    For introductory purposes, let us consider a simplified example involving two pairs of variables $x_1, y_1$ and $x_2, y_2$. We expect some relationship between $y$ and $x$, with noise, such that $y = f(x) + \epsilon$. However, we have no functional form for $f(x)$. To proceed, we will assert that the joint distribution of $y_1$ and $y_2$ is Gaussian with mean 0 and equal variances $\sigma^2$. Now let us consider the conditional probability of $P(y_1 | y_2)$, given that $y_2$ now has some known value, say $y^*$:
    \begin{subequations}
    \begin{align}
        P(y_1 | y_2 = y^*) &\sim \mathcal{N}(\mu_1,\sigma_1) \, \\
        \mu_1 = \rho y^* \ , \\
        \sigma_1 = \sqrt{1-\rho^2} \sigma \ , 
    \end{align}
    \end{subequations}
    where $\rho$ is the correlation coefficient between $y_1$ and $y_2$. Within this toy model, our conditional mean and variance prediction on $y_1$ depends only on the degree of correlation between $y_1$ and $y_2$. Gaussian Processes make a statement on this degree of correlation by proposing that:
    \begin{equation}
        \rho \propto k(x_1,x_2) \ ,
    \end{equation}
    Where $k(x_1,x_2)$ is the kernel function. In other words, GPs say (depending on the choice of kernel) that data points' proximity in $x$ dictates their level of correlation in $y$. In practice, GPs handle arbitrarily many inputs (often $x \in \mathcal{R}^d$) and produce real-valued function outputs. The hyperparameters of the GP kernel, as well as $\sigma$, are found by likelihood maximisation over the observed data, which can be noisy. We direct the interested reader to references which give a full background on GPs \cite{Santner2003,Williams2006}.

    For our purposes, we must make choices on the GP kernel\cite{Duvenaud2014} and then fit the GP to the training data. For single-fidelity settings (all the training data comes exclusively 1D \textit{or} 2D simulations), we used the Matern kernel\cite{Williams2006} with anisotropic lengthscales (i.e. a lengthscale per design parameter):
    \begin{subequations}
    \begin{align}
      k_\nu(\mathbf{x}, \mathbf{x}')
      \;=\;&
      \frac{2^{1-\nu}}{\Gamma(\nu)}
      \left( \sqrt{2\nu}\, \rho \right)^{\nu}
      K_{\nu}\!\left( \sqrt{2\nu}\, \rho \right), \\ 
      \rho \;=\;& \Bigg(\sum_{i=1}^{d} \frac{(x_i - x_i')^2}{\ell_i^2}\Bigg)^{\!1/2}.
    \end{align}
    \end{subequations}
    Where $l_i$ is the $i$-th lengthscale, $\nu$ is the smoothness parameter (we use 5/2), and $K_\nu(x)$ is the modified Bessel function of the second kind of order $\nu$. Fitting to the training data yields best fit values for the lengthscales $l$ and signal variance $\sigma^2$.
    
    However, GPs have some key limitations, a key issue being their poor scaling with training data size. For $\mathcal{O}$(10k) samples, training the GP can become excessively slow. In practice, we found training a GP surrogate with close to the full dataset became slower than performing a Chimera simulation, even when using a GPU for GP training ($\sim$ 10 min training time). This was particularly problematic for multi-fidelity (i.e. using 1D \textit{and} 2D data) GPs where we modified the kernel to introduce discrete fidelity dependence. In these scenarios, we needed an alternative surrogate model which scaled well with training data volume.

    \subsubsection{Ensemble of Neural Networks}\label{section:NN}

    Neural networks (NNs) are universal function approximators and can therefore act as surrogate models, with previous examples already in the field of ICF\cite{Spears2025,Ejaz2024,Gaffney2024}. Similar to GPs, NNs use training data to learn a functional relationship between the inputs ($X$) and the outputs ($Y$). In this work, we used a multi-layer perceptron (MLP) NN architecture which involves a sequence of layers of fully-connected ``neurons", the basic unit of a MLP. Neurons apply a linear transformation to the incoming data before applying a non-linear activation function. Designing a good MLP architecture involves the tuning of a modest number of hyperparameters, for example: number of layers, number of neurons per layer, activation function, learning rate, and regularisation parameters. 
    
    Unlike Gaussian Processes, NNs do not inherently provide any uncertainty estimate in their prediction of $Y$ given $X$. A common method to model uncertainty with NNs is the ensemble approach \cite{Lakshminarayanan2017}. For this, a number of NNs are trained using different random splits of the training data. This causes their predictions to vary, giving a range of results. This is used to quantify the uncertainty in the prediction. To ensure meaningful uncertainty predictions, we performed a calibration scaling on the ensemble predictions, as detailed in \cref{appendix:calibration}.
    
\begin{figure}[htp]
    \centering
    \includegraphics*[width=0.99\columnwidth]{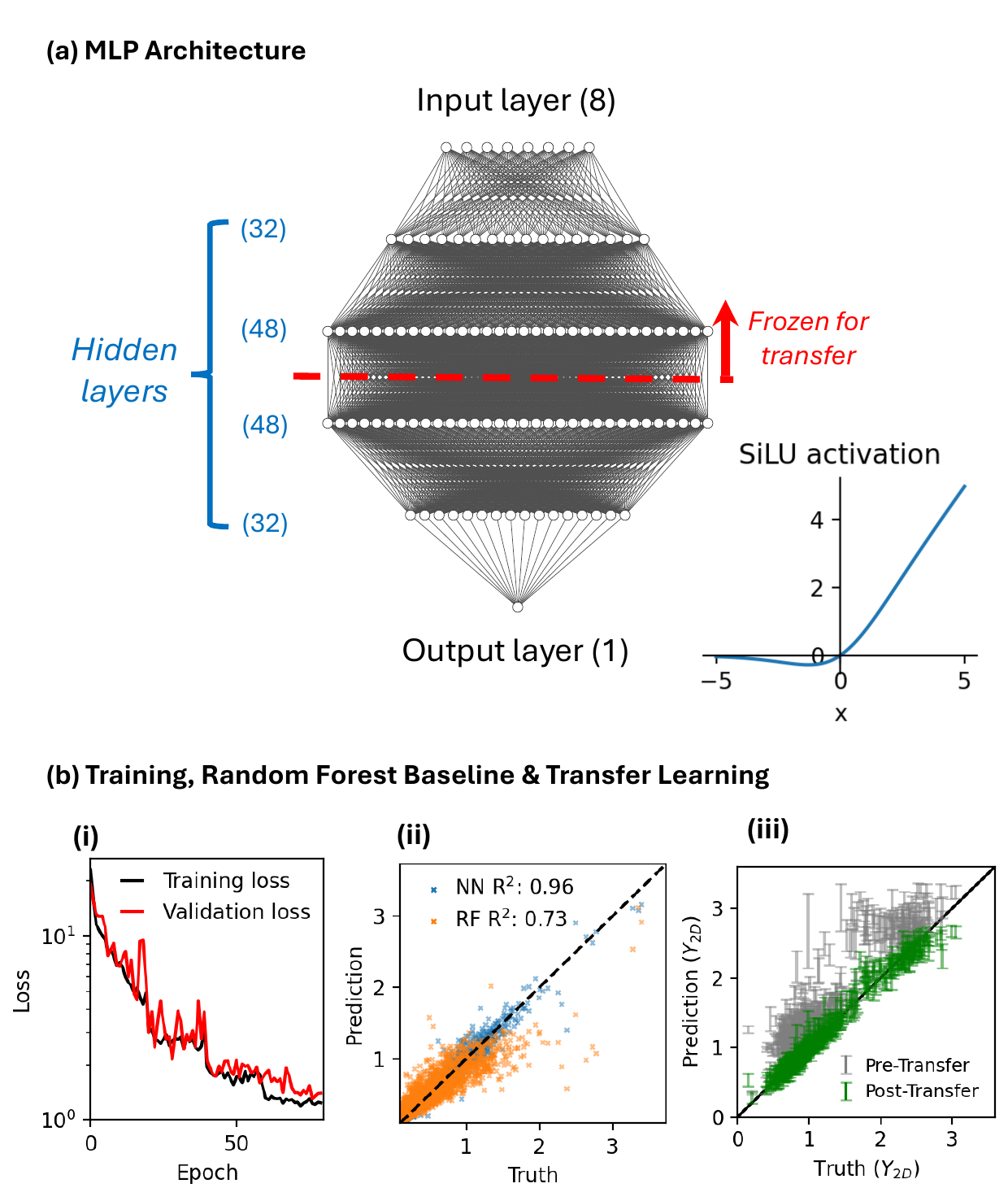}
    \caption{(a) Shows the final architecture chosen for the multi-layer perceptron surrogate model. Each ensemble model consists of 25 instantiations of this model. The diagram shows the number of layers, neurons, activation function and transfer learning freezing strategy. (b) Three figures showing the performance of this surrogate model . (i) Loss history for training of a single 1D surrogate NN. (ii) Comparison on prediction performance between a trained NN surrogate and a random forest (RF)\cite{Breiman2001} regressor on the same validation data. The $R^2$ metric denotes the square of the Pearson correlation coefficient. (iii) Comparison of the ensemble model predictions of the 2D dataset (described later) before and after transfer learning has occurred.}
    \label{fig:NNEnsembleDiagram}
    \end{figure}

    An ensemble size of 25 was used for the 1D and 2D ensemble models. The MLP model was written using PyTorch\cite{Paszke2019}, which allowed its easy integration in the Bayesian optimisation framework BOTorch\cite{Balandat2020}, see \cref{section:BO}. A combination of manual and automated tuning (number of layers, neurons, batch size, learning rate) was used to find the final architecture, shown in \cref{fig:NNEnsembleDiagram}. Training of each member of the ensemble used the AdamW gradient-based optimiser\cite{Loshchilov2017} to minimise the mean squared error between prediction and truth values in the training data. Weight-decay (value of 0.05) was used to regularise the model and prevent over-fitting to the training data. A learning rate schedule (starting at 0.01 and halving every 20 epochs) was found that allowed rapid training in 80-100 epochs (passes over the full training data set) for the 1D surrogate.

    Another key advantage of NN surrogate models is the ease of transfer learning\cite{Humbird2019}. In the context of this work, this means an ensemble of NN surrogates trained on 1D simulation data can be transferred to predict the outcome of 2D simulations, even with limited 2D data. The transfer learning process proceeds as follows:

    \begin{enumerate}
        \item Train an ensemble of NN surrogates on randomised train-test-validation (60:20:20) data splits of the 1D data set
        \item The first N layers of the NN surrogates are `frozen', such that their trained parameters will not change in subsequent training
        \item Perform transfer learning by re-training the ensemble of NN surrogates on randomised train-test (70:30) data splits of the 2D data set
    \end{enumerate}

    In this manner, information from the 1D (correlations between input parameters, useful non-linear embeddings of the inputs, etc.) will be transferred to a 2D surrogate. In practice, it was found the freezing all but the last 2 layers produced good results when tested on validation data.

    \subsection{Active learning}

    The input parameters for laser direct drive (target dimensions, laser pulse parameters) are highly correlated in good designs. For example, shocks launched by the laser are optimal if they can be merged together precisely\cite{Munro2001}, this correlates the thicknesses of target layers to laser parameters like picket powers and durations. This correlation leads to good designs residing within small, complex volumes in the design/input space. To enable effective optimisation, we wish to populate these volumes with more sample data rather than uniformly filling the input space\cite{gopalaswamy2025automated}. We opted for an active learning\cite{Blau2024} approach to achieve this.

    For active learning, we will use a surrogate model trained on uniformly sampled data to narrow down the search space for good designs. Based on the surrogate model predictions, new sample data will be taken (via rad-hydro sims) and used to update the surrogate model. In this way, we can increase the sample frequency within regions we care about, while training the surrogate simultaneously. This also has the additional benefit that batches of candidate points with high objective function values can be pulled from the surrogate model and run concurrently, i.e. in batches. In this work, we employed the active learning algorithm described in \cref{algo:active}. This shares features with canonical level set estimation\cite{Gotovos2013} algorithms.

    Where $\Phi\left(X\right)$ is the cumulative normal distribution and $\mathcal{U}(a,b)$ is the random uniform distribution between $a$ and $b$. There are a number of hyper-parameters to be chosen, $P_{\mathrm{min}}$ and $\hat{Y}_{\mathrm{threshold}}$. It was found that $P_{\mathrm{min}}$ values between 0.25 and 0.5 gave a good balance between exploration and exploitation. The threshold is chosen such that designs with expected outcomes better than the threshold are searched for.
    
    \begin{algorithm}[H]
    \caption{Active Learning: Probabilistic Threshold Sampling}\label{algo:active}
    \begin{algorithmic}
    \Require Number of learning runs $N_{\mathrm{runs}}$, batch size $N_{\mathrm{batch}}$, retrain interval $r$, threshold $\hat{Y}_{\mathrm{threshold}}$

    \For{$i = 1$ to $N_{\mathrm{runs}}$}
        \If{$i \cdot N_{\mathrm{batch}} \bmod r = 0$}
            \State Retrain surrogate model on $X, Y$
        \EndIf

        \State $M \gets 0$
        \While{$M < N_{\mathrm{batch}}$}
            \State Generate $s$ random draws $u_j \sim \mathcal{U}(P_{\mathrm{min}}, 1)$
            \State Generate $s$ random input points $X_j$
            \State Predict mean $\mu_j$ and std $\sigma_j$ from surrogate
            \State Compute $P(y_j > \hat{Y}_{\mathrm{threshold}}) = 1 - \Phi\left(\frac{\hat{Y}_{\mathrm{threshold}} - \mu_j}{\sigma_j}\right)$
            \State Select $X_{\mathrm{cand.}}$ from $X_j$ if $P_j > u_j$
            \State $M \leftarrow$ number of points $X_{\mathrm{cand.}}$
            \State If $M < N_{\mathrm{batch}}$: Double random sample size $s$
        \EndWhile
    
        \State Select $N_{\mathrm{batch}}$ inputs ($X_{\mathrm{next}}$) at random from $X_{\mathrm{cand.}}$
        \State $(P_{\mathrm{next}}, Y_{\mathrm{next}}) \gets \operatorname{Chimera}(X_{\mathrm{next}})$
        \State Add $X_{\mathrm{next}}, P_{\mathrm{next}}, Y_{\mathrm{next}}$ to dataset

    \EndFor
    \end{algorithmic}
    \end{algorithm}

    \subsection{Bayesian optimisation}\label{section:BO}

    Initial sampling and active learning was used to build a comprehensive data set to train a sufficiently accurate surrogate model to be used in Bayesian optimisation. Bayesian optimisation defines an optimisation problem in terms of maximising an acquisition function, which itself is a function of the surrogate model mean and uncertainty predictions, $\mu(X)$ and $\sigma(X)$ respectively. This optimisation problem over the acquisition function is computationally cheaper than the `true' optimisation problem over the simulator outputs.
    
    We used the same Bayesian optimisation strategy throughout this work. Firstly, we restrict to simulation batch sizes of 1 i.e. sequential optimisation. This allows for the most well defined acquisition function and optimisation problem. Secondly, we used the log Expected Improvement\cite{Ament2023} acquisition function, which is defined as:

    \begin{subequations}
    \begin{align}
        \mathrm{LEI}(X) &= \log\left[\sigma(X)\left(Z\Phi(Z)+\phi(Z)\right)\right] \, \\
        Z &= \frac{\mu(X)-\mathrm{max}(Y)}{\sigma(X)}
    \end{align}
    \end{subequations}

    Where $\Phi\left(X\right)$ and $\phi\left(X\right)$ are the cumulative and probability distribution functions of the standard normal distribution. This acquisition function is maximised w.r.t. $X$ using gradient-based methods. A new simulation is then ran at the optimal $X$ found. This produces a new $X, Y$ pair for the dataset, the surrogate is retrained and the process is repeated. The numerical solution to this optimisation problem was performed by the BOTorch library\cite{Balandat2020} and the coupling to the simulator is handled by software described in the following section.

    It should be noted that we have not taken a fully multi-fidelity Bayesian optimisation approach\cite{Folch2023}, even if we are using multi-fidelity surrogates. This is because of the large difference in computational cost ($\sim$ 200x) between our fidelities (1D or 2D rad-hydro sims). In traditional multi-fidelity Bayesian optimisation, the optimiser chooses the next candidate $X$ \textit{as well as} the fidelity based on the information gain per unit cost. Due to our cost disparity, 2Ds are run very sparingly so cost weighting would need to be used. Instead, we opted to run Bayesian optimisation at a fixed fidelity but using surrogates trained on both fidelities. This allows the transfer of information between fidelities via the surrogate but updates to the database are strictly within a single fidelity.

    \subsection{Orchestration}

\begin{figure}[htp]
    \centering
    \includegraphics*[width=0.99\columnwidth]{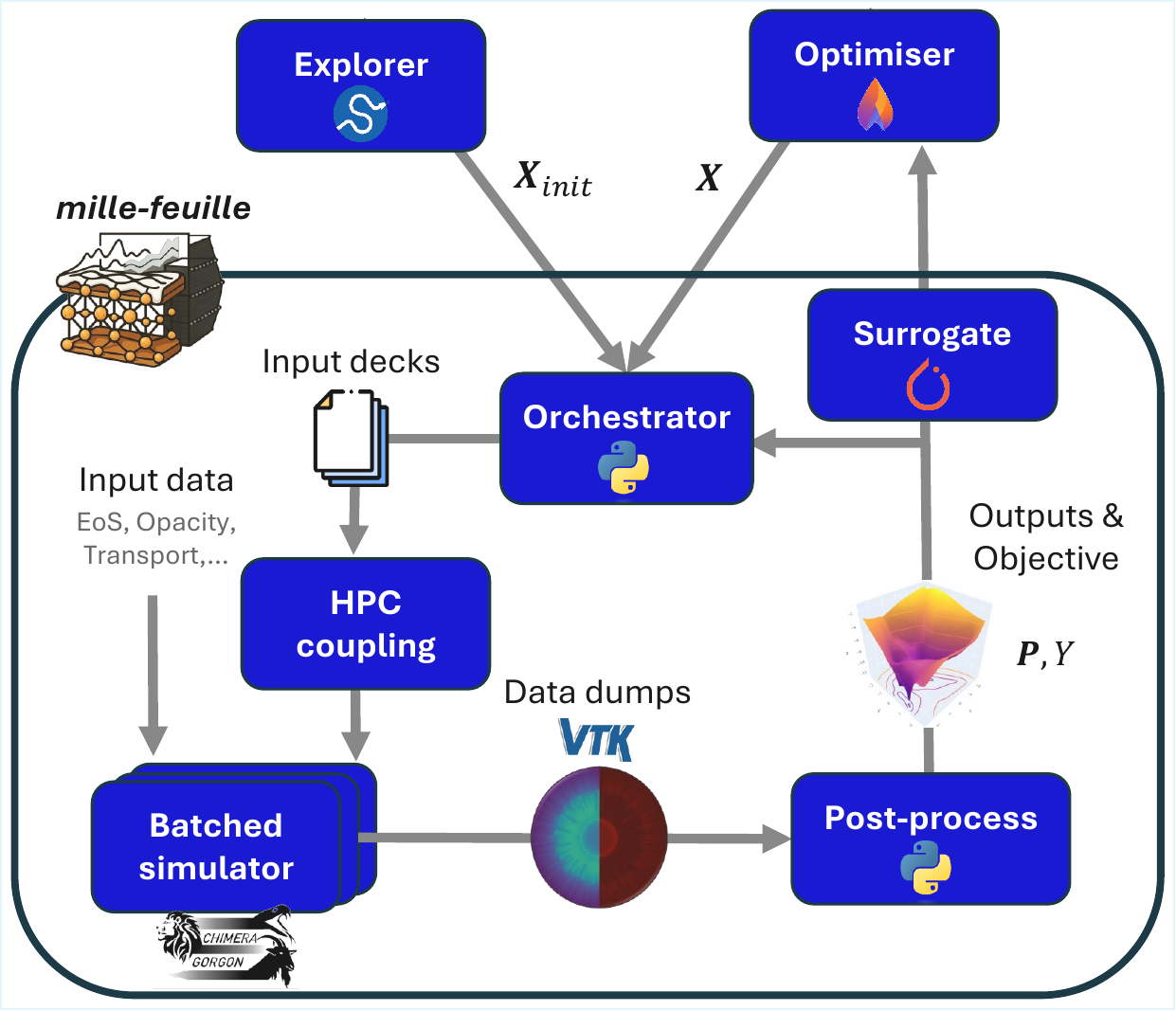}
    \caption{A schematic showing the ML and simulation workflow used in this work. The orchestration software, \textit{mille-feuille}, ensures the new inputs suggested by random sampling or an optimiser are correctly converted to input decks for the simulator. It also handles the coupling of the training database and the surrogate used in optimisation.}
    \label{fig:millefeuille}
    \end{figure}

    An orchestration framework is required to efficiently couple together the machine learning and simulator models described above. The job of orchestration software is to: convert candidate sample points into valid simulator inputs; execute and wait for batches of simulations to run within a HPC environment; post-process the simulations for valid training data; re-train the surrogate model and finally query the explorer/optimiser for new sample points. \cref{fig:millefeuille} summarises this workflow. 
    While the specifics of this workflow are simulator and surrogate model dependent, the software package developed for this purpose, \textit{mille-feuille}, is open-source\cite{Crilly_millefeuille} and aims to abstract this workflow process.

    Creating automated design programs in this way allows multiple levels of parallelism to be exploited. At the lowest level, the simulator is parallelised by MPI with load balanced domain decomposition. Within the automated design process, batches of the simulator can be launched simultaneously to use all the allocated HPC resource. At the top level, multiple instances of the automated design program can be launched as an array, learning from the same centralised data sets.
    
    \section{Optimisation Study}

    \subsection{1D design}

\begin{figure}[htp]
    \centering
    \includegraphics*[width=0.99\columnwidth]{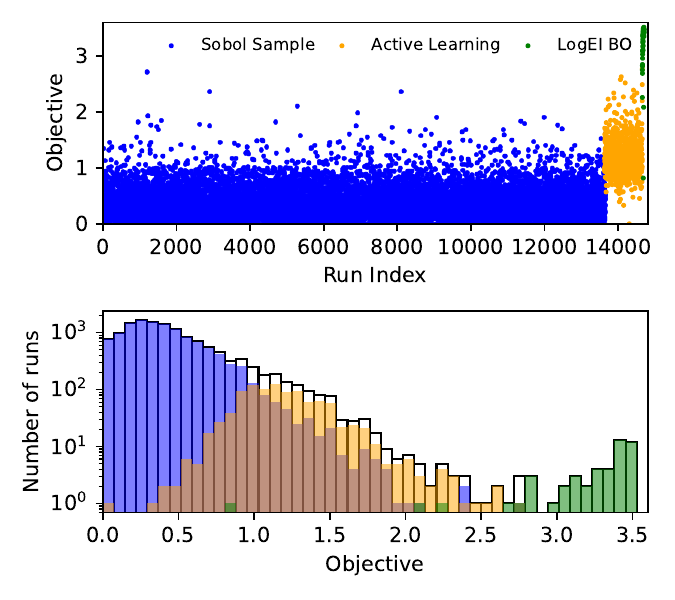}
    \caption{(Top) History of objective values obtained from 1D simulations, points are coloured by which sampling/optimisation method created them. (Bottom) A histogram of objective values over the full 1D database, note the log y scale.}
    \label{fig:1DObjectiveHist}
    \end{figure}

    An extensive initial sampling 1D simulation ensemble was run due to the relatively low cost of 1D simulations. The input domain defined in \cref{tab:design} was transformed to the unit-hypercube and the quasi-random Sobol sequence\cite{Sobol1967} was used as a space filling design\cite{Santner2003}. Of order 12.5k 1D simulations were ran based on this Sobol design. As expected the vast majority (97\%) of these simulations represented poor designs (quantified as objective $Y < 1$) which are unlikely to ignite at the 2MJ scale. However, of those simulations with $Y > 1$ the range of the input parameters still covered the whole design space. This is because the inputs for this dataset of ``good'' designs are correlated. For example, ice and ablator thickness are anti-correlated with each other (-0.6 Pearson coefficient) and outer radius (-0.16 and -0.35 respectively), suggesting good designs lie within a certain target mass range. For the laser pulse, picket power is positively correlated with foot power (+0.35) and negatively with the delay between picket and foot (-0.3). This reflects the importance of the front end of the pulse in shock timing \cite{Munro2001}. The distribution of objective values from this initial sampling can be seen in \cref{fig:1DObjectiveHist}.

    The comprehensive initial sampling allowed the training of a single fidelity GP surrogate for active learning. Since only 3\% of the initial sampling lay within the region of interest ($Y > 1$), active learning with a threshold of 1 and $P_{min}$ = 0.25 was used to find $\sim$ 1k points within this region of interest. As expected, the active learning sample shows similar statistical correlations as the thresholded initial Sobol samples. The active learning step increased the population of high objective function observations, increasing surrogate model confidence in this region.

    Finally, Bayesian optimisation was performed to find the optimal design in 1D within the provided design space. Two approaches were taken. Firstly, a GP surrogate model was used, however training time was becoming obstructive due to the large 1D dataset. For the GP, the training data was downselected by reducing the design space to encompass all points with $Y > 1.5$, this reduced the number of training samples from $\sim$ 14k to $\sim$ 5k. Secondly, the NN ensemble was used, for which the whole 1D dataset was kept. Both surrogates were used to optimise Log Expected Improvement. All but the last two layers of the NNs were frozen for retraining during the optimisation loop. The optimisations with the different surrogates converged towards similar optimal designs ($Y$ values within 0.1 of each other), however it was the NN surrogate which found the highest performing design after $\mathcal{O}(10)$ new samples.

    \subsection{2D design}

    It is evident that any target optimisation using only 1D simulations will produce optimistic designs, as the designs do not need to be robust to hydrodynamic instabilities to achieve high performance in 1D. However, there are great benefits to transferring some of the information learned from 1D simulation ensembles to optimisation using multi-dimensional radiation hydrodynamics simulation. In particular, 1D results represent the upper limit of performance for 2D calculations. Therefore, by transferring information about the design space from 1D to 2D we can reduce the search space drastically.

    \begin{figure}[htp]
    \centering
    \includegraphics*[width=0.95\columnwidth]{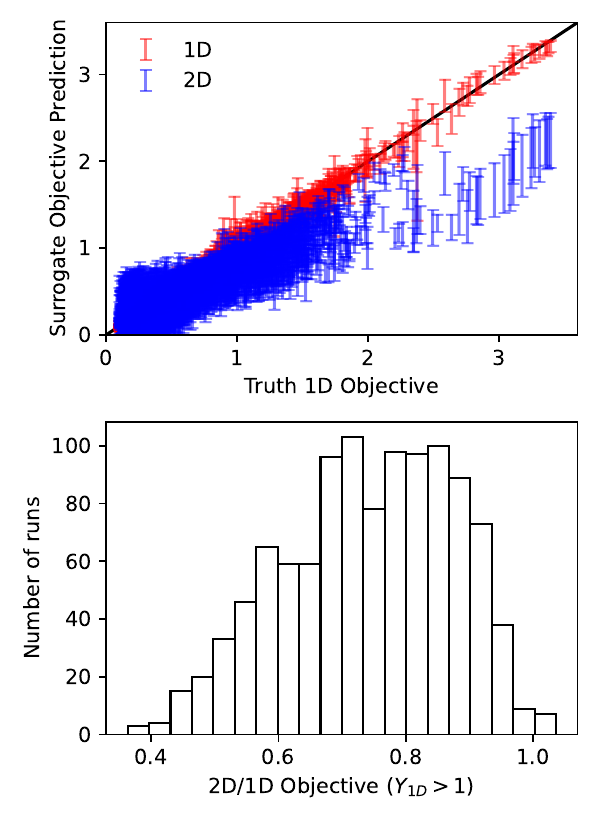}
    \caption{Trained 1D and 2D neural network ensemble surrogate model predictions of the objective function for input values from the 1D database. Transfer learning for the 2D surrogate only makes use of the transferred samples as a training set.}
    \label{fig:EnsemblePredictions}
    \end{figure}

    Therefore, our 2D design process proceeded as follows. A subset of the inputs from 1D active learning samples were re-run as 2D simulations, we will refer to this as our ``transfer sample". This transfer sample allowed the training of 2D surrogate models. While both multi-fidelity GPs and NN ensembles were tested, it was found that the NN ensembles performed equally well with considerably lower training times. Trained on the 1D dataset and the 2D transfer samples, surrogate model predictions for both 1D and 2D were compared for inputs ($X$ values) in the 1D dataset. In \cref{fig:EnsemblePredictions}, we can see the 2D surrogate has learnt from the transfer sample that performance is degraded relative to 1D. It is also observed that the degradation is a strong function of input parameters, particularly ice thickness for these simulations. It is worth repeating that the 2D data also is inherently noisy as different random surface perturbations are applied for each simulation. This is reflected in the large surrogate uncertainty shown in the error bars, noting that these uncertainties have been calibrated using the method shown in \cref{appendix:calibration}.

\begin{figure}[htp]
    \centering
    \includegraphics*[width=0.99\columnwidth]{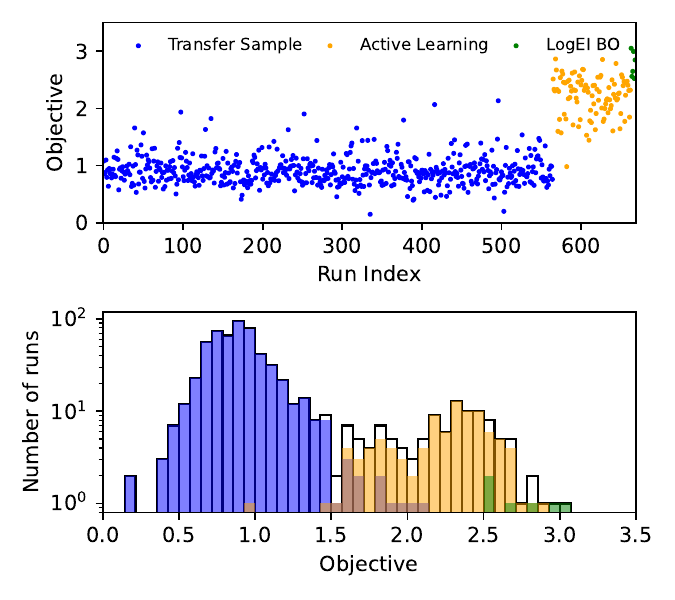}
    \caption{(Top) History of objective values obtained from 2D simulations, points are coloured by which sampling/optimisation method created them. (Bottom) A histogram of objective values over the full 2D database, note the log y scale.}
    \label{fig:2DObjectiveHist}
    \end{figure}

    Approximately 128 2D simulations were run within an active learning loop to explore the space of good 2D designs. For the active learning, a threshold objective value of $2$ and $P_{min}$ of 0.5 were used, aiming to exploit the space rather than to explore. \cref{fig:2DObjectiveHist} shows the new samples created by the active learning method. This clearly shows the transfer learning has been effective in narrowing down the search space for 2D designs based on the 1D dataset. The 2D surrogate directs samples towards large objective values and shows no examples of poor design choices (lowest objective value $\sim 0.9$). The transfer sample does not contain poor 1D designs, $Y < 1$, so it is clear that the understanding of the design space found in 1D is maintained in the transfer learning, in particular what regions of design space to avoid due to low performance.
    
    Finally, Bayesian optimisation was performed using the 2D NN ensemble surrogate only, with transfer learning performed using all available 1D and 2D data. With a small number of runs, the algorithm found a small region of input space with robust 2D performance, with a maximal objective function of approximately $3.0$, to be compared to $3.5$ in 1D.

    \section{Results}

    In the following two sections we will provide a detailed examination of the performance of the optimal designs at both the design scale (25 kJ) and scaled up (2 MJ). \cref{tab:performance} summarises the key performance parameters for an overview of the design performance.

    \begin{table*}[t]
        \centering
        \begin{tabular}{c|c|c|c|c|c|c|c|c|c}
        Scale & 1D/2D & Design & $\chi_{S, no \ \alpha}$ & $Y$ & DT n Yield & $\langle \rho$R$\rangle$ (g/cm$^2$) & $\langle$T$_i\rangle$ (keV) & $v_{imp}$ (km/s) & Bang Time (ns) \\
    \hline
        25 kJ & 1D & $O_{1D}$ & 2.0 & 3.52 & 5.49 $\times$ 10$^{14}$ & 0.370 & 4.35 & 430 & 3.67 \\
          &  & $O_{2D}$ & 1.85 & 3.40 & 4.10 $\times$ 10$^{14}$ & 0.378 & 4.16 & 416 & 3.50 \\
          & 2D & $O_{1D}$ & 1.65 & 2.02 & 2.29 $\times$ 10$^{14}$ & 0.266 & 4.22 & 434 & 3.67 \\
          &  & $O_{2D}$ & 1.72 & 3.05 & 3.50 $\times$ 10$^{14}$ & 0.348 & 4.08 & 416 & 3.50 \\
    \hline
        2 MJ (Burn Off) & 2D & $O_{1D}$ & - & - & 5.86 $\times$ 10$^{16}$ & 1.20 & 4.84 & 433 & 15.80\\
          &  & $O_{2D}$ & - & - & 1.02 $\times$ 10$^{17}$ & 1.46 & 4.85 & 416 & 15.05\\
    \hline
        2 MJ (Burn On) & 2D & $O_{1D}$ & - & - & 1.02 $\times$ 10$^{18}$ & 1.00 & 7.89 & 433 & 15.92 \\
          &  & $O_{2D}$ & - & - & 2.24 $\times$ 10$^{19}$ & 1.16 & 18.18 & 416 & 15.12 \\
        \end{tabular}
        \caption{Performance of the optimal designs at different laser energy scales. Note that full hydrodynamic equivalency re-tuning was not performed when hydro scaled to 2 MJ. Burn off/on refers to simulations excluding/including alpha particle heating.}
        \label{tab:performance}
    \end{table*}

    \subsection{Performance of optimal designs at 25 kJ scale}

    We will first consider the result of the 1D and 2D design optimisation at the design energy scale of 25 kJ. We will assess the design performance using the objective function, \cref{eqn:objective}, as well as key ICF performance parameters\cite{Lindl2018} (DT fusion yield, ion temperature, implosion velocity, etc.). The optimal design parameters are given in \cref{tab:optdesign} and the key performance parameters are summarised in \cref{tab:performance}.

    \begin{table}[h]
        \centering
        \begin{tabular}{c|c|c}
        Variable (units) & $O_{1D}$ & $O_{2D}$ \\
        \hline
             $P_{\mathrm{picket}}$ (TW) & 3.19 & 4.53 \\
             $P_{\mathrm{foot}}$ (TW) & 0.65 & 0.80 \\
             $t_{\mathrm{delay}}$ (ps) & 806 & 757\\
             $t_{\mathrm{foot}}$ (ps) & 782 & 668\\
             $t_{\mathrm{rise}}$ (ps) & 1000 & 995\\
             $R_{\mathrm{outer}}$ (um) & 436 & 429 \\
             $\Delta_{\mathrm{ablator}}$ (um) & 9.5 & 9.5 \\
             $\Delta_{\mathrm{ice}}$ (um) & 70.5 & 76.0 \\
        \end{tabular}
        \caption{Optimal designs at the 25 kJ scale derived from 1D ($O_{1D}$) and 2D ($O_{2D}$) simulation based optimisation.}
        \label{tab:optdesign}
    \end{table}

\begin{figure}[htp]
    \centering
    \includegraphics*[width=0.94\columnwidth]{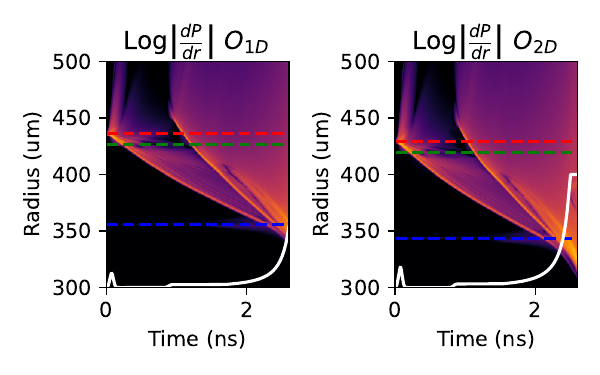}
    \caption{Heatmaps of the logarithmic pressure gradient against radius and time (in the Eulerian frame). The initial ablator radius, ablator-ice interface and ice-gas interface are shown with red, green and blue dashed lines respectively. Shown in white is the incident laser power (arbitrary scale).}
    \label{fig:1DShocks}
    \end{figure}

    The relative differences between the 1D and 2D optimal designs (denoted $O_{1D}$ and $O_{2D}$ respectively) give rise to only small differences in performance in 1D, with areal densities within 2\% of each other, with the $O_{1D}$ design having a higher $\chi_{S, \mathrm{no} \alpha}$ value. To understand these differences in the designs we will look at important implosion characteristics related to both 1D performance and hydrodynamic stability. By inspecting the pressure gradient shown in \cref{fig:1DShocks},
    we can see the optimiser achieves a reasonable degree of shock timing for both designs. It is worth noting that this is not specifically optimised for, but arises when maximising our chosen objective. This allows these designs to achieve high areal densities (370 - 380 mg/cm$^2$). These areal densities are considerably higher than has been achieved in experiment at OMEGA\cite{Regan2012} ($\lesssim$ 250 mg/cm$^2$), pointing to the optimism of 1D design calculations.

\begin{figure}[htp]
    \centering
    \includegraphics*[width=0.9\columnwidth]{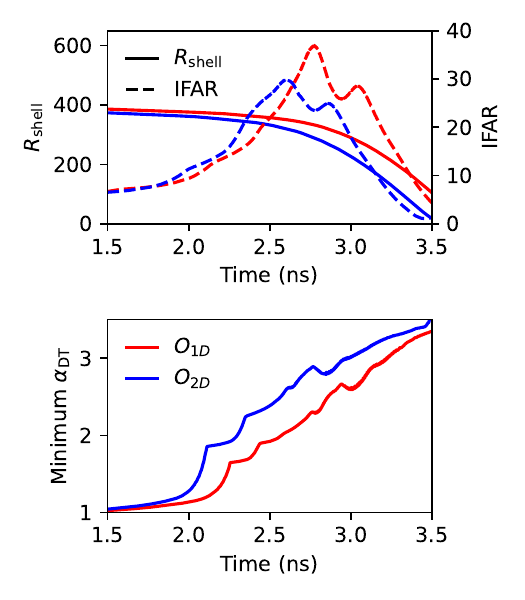}
    \caption{(Top) Time series of the inflight aspect ratio (IFAR) and shell radius in dashed and solid lines respectively. (Bottom) Time series of the minimum fuel adiabat for the two optimised designs.}
    \label{fig:InflightStability}
    \end{figure}

    From the 1D simulations, we can also look at key metrics of hydrodynamic stability. These are the inflight aspect ratio\cite{mckenna2013} (or IFAR) and the minimum adiabat\cite{lindl1995} of the DT fuel, $\alpha_{DT}$. These stability metrics are plotted as a function of time in \cref{fig:InflightStability}. 
    Higher IFAR means the imploding shell is thin in-flight and therefore more easily disrupted by the non-linear growth of Rayleigh-Taylor. The $O_{1D}$ and $O_{2D}$ have peak IFARs of 37 and 30 respectively. Another key stability metric is the shell adiabat (defined the ratio of the total pressure to the degeneracy pressure). Low adiabat shells are more compressible leading to higher final densities at stagnation. However, low adiabats are unfavourable for hydrodynamic stability\cite{Takabe1985,Betti1998}. \cref{fig:InflightStability} shows that $O_{2D}$ has a consistently higher adiabat than $O_{1D}$. This is achieved by using a higher picket power (4.5 TW vs 3.2 TW), as well as a degree of shock mistiming seen in \cref{fig:1DShocks}.

    \begin{figure}[htp]
    \centering
    \includegraphics*[width=1.0\columnwidth]{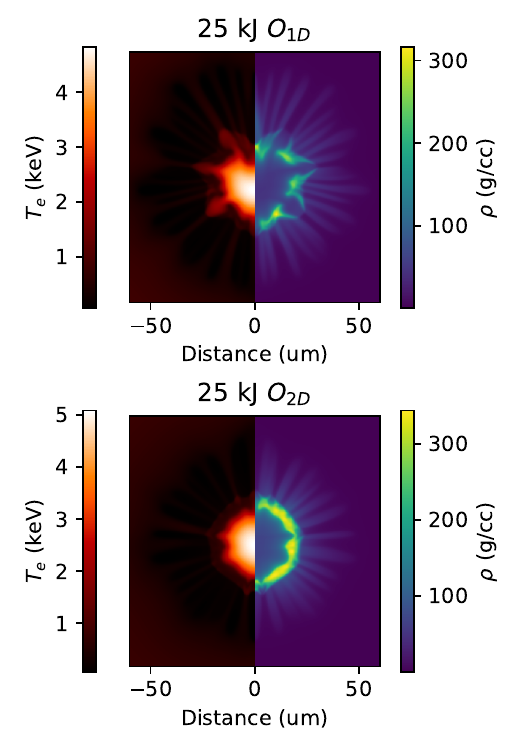}
    \caption{Bang time conditions (mass density $\rho$ and electron temperature $T_e$) from 2D Chimera simulations of the optimised 1D and 2D designs (top and bottom respectively).}
    \label{fig:2DBangTimeOMEGA}
    \end{figure}

    As expected from analysing the stability in 1D, multi-dimensional Chimera simulations show that the $O_{1D}$ design degrades far more than the $O_{2D}$ design in the presence of hydrodynamic instabilities. This is particularly evident in the bangtime density and temperature distribution as shown in \cref{fig:2DBangTimeOMEGA}. The DT shell in the $O_{1D}$ design is destroyed by the growth of hydrodynamic instabilities, whereas the $O_{2D}$ design keeps the shell relatively intact. The OMEGA beam mode (averaged into 2D $r,\theta$ geometry) is particularly evident in the $O_{1D}$ fuel morphology. It is clear the $O_{2D}$ design (while similar in many design parameters to $O_{1D}$) is more robust to hydrodynamic instabilities.

\begin{figure}[htp]
    \centering
    \includegraphics*[width=0.99\columnwidth]{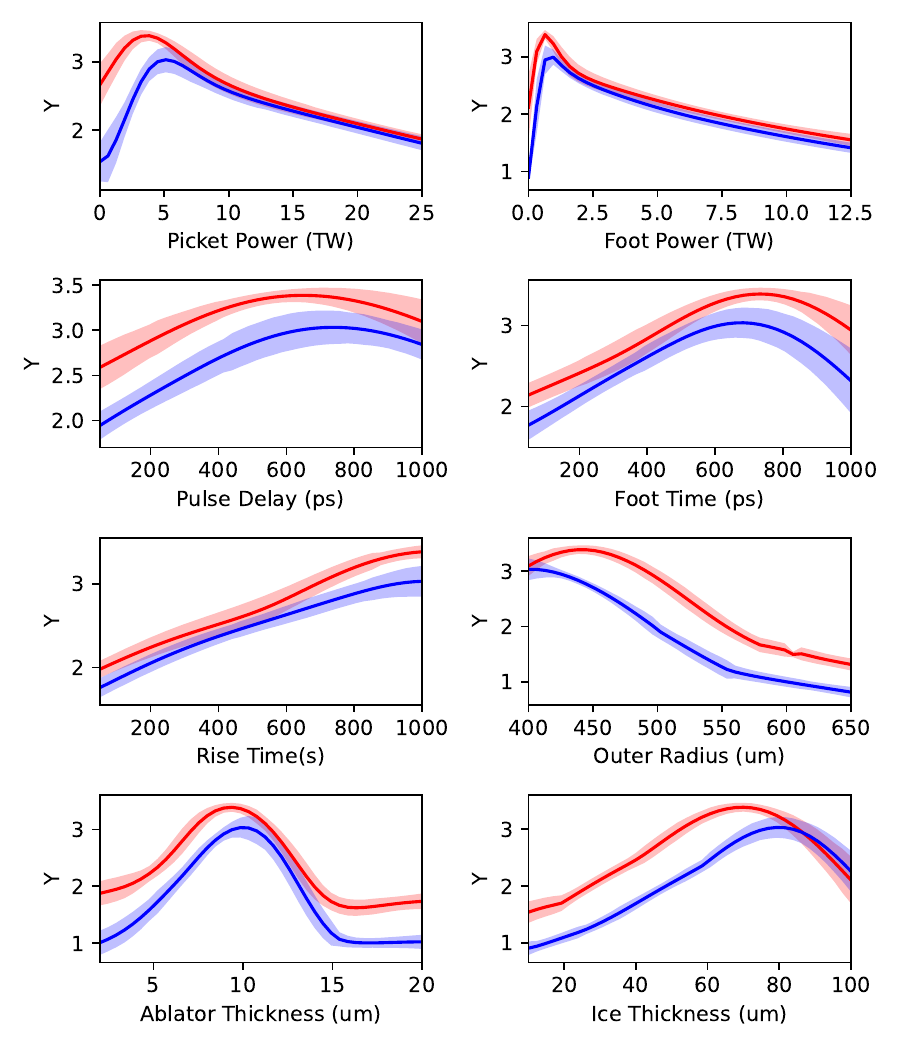}
    \caption{Neural network ensemble surrogate model predictions of the optimal performance at a given value of design parameter. Each subplot shows the variation of 1 of the 8 design parameters, where the other 7 design parameters have been optimised over. The 1D and 2D surrogate predictions are shown in red and blue respectively. Calibrated error bars are shown.}
    \label{fig:SurrogatePredictions}
    \end{figure}

    The trained 1D and 2D surrogate models are also a useful tool to visualise and understand the effect of a single design decision. We can optimise over the surrogate model's (mean) prediction of the objective, while holding one input parameter at a fixed value. The results from this investigation are shown in \cref{fig:SurrogatePredictions}. These results show some intuitive findings when comparing the predictions for 1D and 2D. For example, in 2D decreasing picket power below a threshold rapidly decreases performance. Low picket power creates a low adiabat implosion which is more prone to hydrodynamic instability growth. In a similar vein, thinner ice is more disrupted by the growth of ablator surface perturbations - the key source of degradation in our 2D simulations. Therefore, we can see from these trends that the 2D surrogate has found regions of input space which are robust to the level hydrodynamic instabilities that was seeded. Amplifying the level of ablator density perturbation applied would push the trends further, heading towards smaller capsules, with thicker ice and higher picket power/adiabat.
    
    \subsection{Performance of optimal designs at 2 MJ scale}

    In this work, the design objective was devised such that the optimal targets hydrodynamically scaled to a 2 MJ laser driver would ignite and produce high gain. In particular, via a multi-fidelity approach, we aimed to find targets that were robust to hydrodynamic instabilities seeded at the ablator surface. In this section, we investigate the scaled design performance via direct simulation at the 2 MJ scale with alpha heating physics included. We used the particle Monte Carlo burn package in Chimera as described in Tong \textit{et al.}\cite{Tong2019}. This model was extended to include the additional effect of fuel depletion via a burn fraction advected around with the DT material which was used to reduce the reaction rate.

    Hydrodynamic scaling by scale factor $S$ involves the following dimensional scalings\cite{Nora2014}:
    \begin{align*}
        R &\rightarrow S \times R \ , \\
        t &\rightarrow S \times t \ , \\
        E &\rightarrow S^3 \times E \ ,
    \end{align*}
    Where $R$, $t$ and $E$ are the spatial, temporal and energetic scales. However, due to non-hydrodynamic physics (e.g. radiation and thermal transport) naively applying this scaling to initial conditions does not lead to hydrodynamically equivalent implosions. Nora \textit{et al.}\cite{Nora2014} note additional changes to the laser pulse and target dimensions are required to restore hydrodynamic equivalency, mainly exchange of ablator mass for DT ice. Here we will not perform this re-optimisation but instead apply hydro-scaling to the 2D simulations at a restart time approaching stagnation, making the assumption that a hydro-equivalent implosion can be achieved up to this time.

    First, both designs have sufficient ignition margin to ignite in 2D at the 2 MJ energy scale. However, the yield amplification due to alpha heating (defined as the ratio of neutron yields burn-on to burn-off) are very different with values of 17 and 217 for the $O_{1D}$ and $O_{2D}$ designs respectively. From the burn-off areal densities, one might expect ignited designs to achieve burn fractions of 16\% and 19\% respectively, based on \cref{eqn:Fraley}. However in the burn-on simulations, burn fractions of 0.9\% and 15\% were achieved. This shows canonical 1D burn propagation was not achieved in the $O_{1D}$ design due to large asymmetries in the confining shell. 
    
    \begin{figure}[htp]
    \centering
    \includegraphics*[width=0.99\columnwidth]{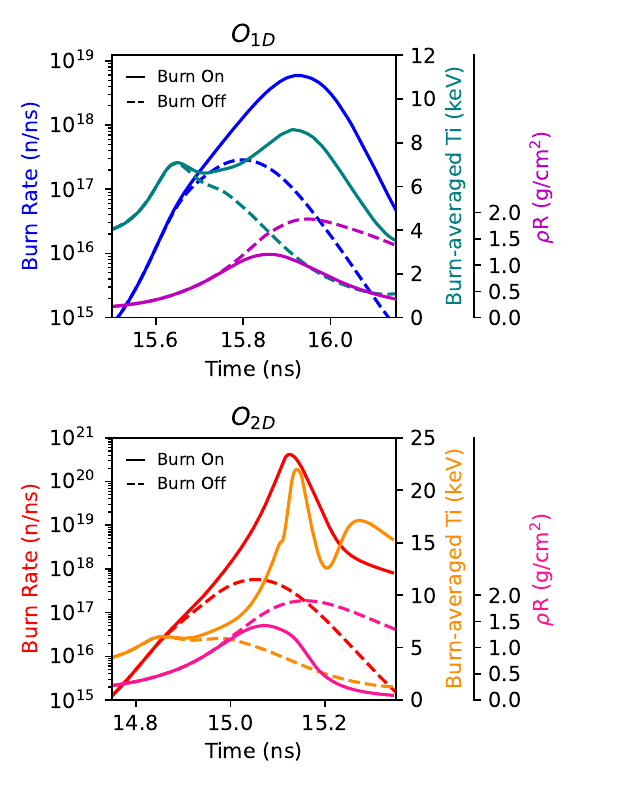}
    \caption{Time series of key burn parameters: burn rate (DT fusion reactions per unit time), burn-averaged ion temperature and areal density (arithmetic average in solid angle). The $O_{1D}$ and $O_{2D}$ design results are shown in the top and bottom subplots respectively. Burn-on and -off are indicated by solid and dashed lines.}
    \label{fig:TimeHistoriesNIF}
    \end{figure}
    \cref{fig:TimeHistoriesNIF} shows the temporal evolution of the burn characteristics for the two designs. Comparing burn-off and -on results for the same design shows higher burn rates and temperatures but lower areal densities. The lower areal densities are due to the rapid expansion driven by increased hotspot pressure from alpha heating. Now contrasting the two designs, the strongly burning $O_{2D}$ design shows super-exponential growth the burn rate, shown by the solid red curve which has positive second derivative in the rise on the log-linear scale. Burn-averaged ion temperature also grows rapidly during this time up to a peak of 22 keV. During the fall of the burn ($t > 15.1$ ns), burn-averaged ion temperature shows a second increase as shock heating of the remaining low density, free-falling shell\cite{Betti2002} produces large ion temperatures (10s keV). This resembles the hohlraum re-heating\cite{Rubery2024} phenomena observed on the National Ignition Facility for ignition experiments. However this free-falling DT heating does not contribute significantly to increased burn. The $O_{1D}$ burn-on simulation shows less pronounced increases in ion temperature and burn rate reflecting the truncated burn propagation. This can be attributed to poor hotspot confinement from thin regions in the shell. This allows the hotspot to form aneurysms through holes in the shell\cite{Hurricane2016,Springer2018}.

    \begin{figure*}[t]
    \centering
    \includegraphics*[width=0.99\textwidth]{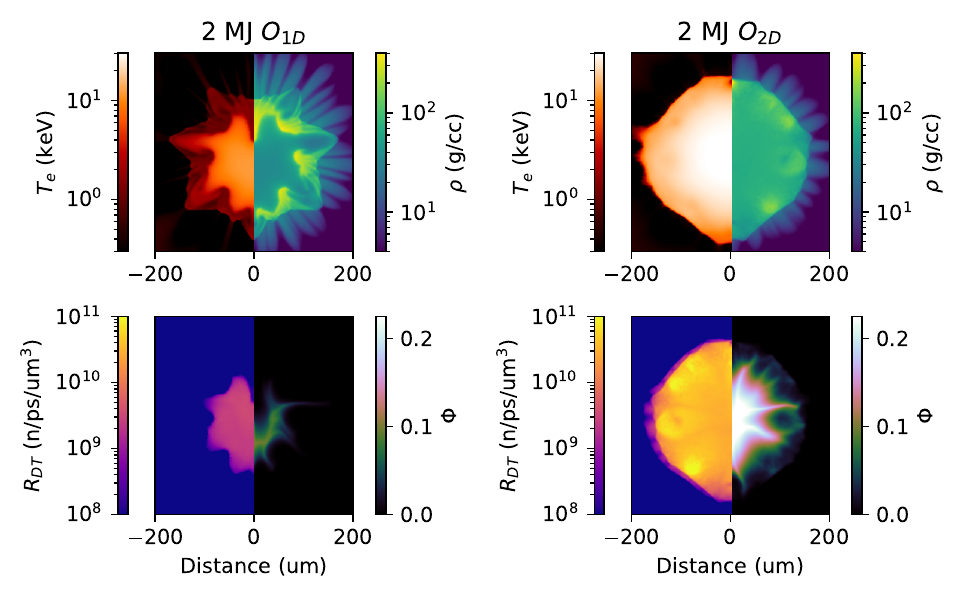}
    \caption{Bang time conditions (mass density $\rho$, electron temperature $T_e$, volumetric DT reaction rate $R_{DT}$ and burn up fraction $\Phi$) from 2D burn-on Chimera simulations at 2 MJ energy scale of the optimised 1D and 2D designs (left and right columns respectively).}
    \label{fig:2DBangTimeNIF}
    \end{figure*}

    The differing degree of burn propagation leads to very different conditions at the time of peak neutron production, as shown in \cref{fig:2DBangTimeNIF}. The strongly burning $O_{2D}$ design has become a near uniform ball of burning and expanding DT. High localised burn up fraction, $\Phi \sim 20\%$, leads to reduced reaction rate, $R_{DT}$, in the core of the burning plasma, where temperatures exceed 30 keV. In contrast, the $O_{1D}$ design shows a largely perturbed shell into which burn has not propagated. Hotspot fuel has blown out through holes in the shell showing it has lost confinement before burn of the dense fuel could be established. This results in the $O_{1D}$ design having far lower performance than the $O_{2D}$ design, as discussed above.

    It is clear from these direct numerical simulations at the 2 MJ scale that the objective function used (\cref{eqn:objective}) and the multi-fidelity methodology was able to design high gain designs at a higher energy scale than simulated. The optimised 2D design allowed for a more robust burn due to its better stability leading to a more uniformly confining shell.

    \section{Future work}

    In this paper we describe a framework for performing multi-fidelity optimisation around the Chimera simulation code. This provides many avenues for future work, using Chimera's full capabilities as a radiation magnetohydrodynamics code. For example, additional fidelities can be introduced such as 3D simulation\cite{Chittenden2016}, the modelling of cross-beam energy transfer\cite{Moloney2024} or the use of experimental data. It is also important to introduce further sources of perturbations, such as DT ice roughness\cite{Nikroo2004,Huang2018} or capsule stalk\cite{Gatu2020}, to ensure optimised designs are robust to instabilities derived from experimentally realistic perturbations. There are also magnetically driven inertial fusion schemes (such as MagLIF) for which optimal designs at higher driver energies are an active area of research\cite{Ruiz2023,Alexander2025}. 

    We have operated with a relatively small design space of 8 parameters. Scaling to larger design spaces is necessary to explore a greater variety of possible designs. Some key additional parameters being considered for next generation inertial fusion designs relate to target layers (dopants, foams)\cite{Maclaren2021,Paddock2023} and laser parameters (beam size, zooming, bandwidth)\cite{Froula2025}. However, the curse of dimensionality means that larger design spaces require larger numbers of sample points to train the surrogate. Additionally, the global optimisation of the acquisition function becomes increasingly difficult in higher dimensions. This will require new optimisation strategies, such as the splitting into smaller sub-problems\cite{gopalaswamy2025automated} or using gradient-based optimisation methods directly from the simulator\cite{Antonova2023,Joglekar2024}.

    \section{Conclusions}

    In this work we show that machine learning models can use simulation data from multiple fidelities to optimise designs in the context of laser direct drive inertial fusion. In particular, we show that a large number of 1D simulations, with inherently no hydrodynamic instability growth, could be used to narrow down the search space for 2D simulations. It was found that an ensemble of neural network based surrogate models was effective at transferring this information between fidelities. This surrogate model also enabled active learning and Bayesian optimisation methods to automatically tune the laser and target design parameters. These methods were robust to noise introduced to the 2D simulation results via randomised ablator density perturbations. 

    The optimal 2D design was more hydrodynamically stable than the optimal 1D design and showed only weak degradation in performance at the 25 kJ energy scale. Hydrodynamically scaled simulations for a 2 MJ laser driver including alpha heating physics showed large differences in the achieved yield amplification, 17 and 217 for the $O_{1D}$ and $O_{2D}$ designs respectively. The $O_{1D}$ design failed to propagate burn into the dense fuel due to loss of confinement from hydrodynamic instability growth. In contrast, $O_{2D}$ design was highly performant at the 2 MJ scale. This demonstrates that our devised objective function and multi-fidelity design framework correctly identified a high-gain scaled-up design, robust to the presence of hydrodynamic instabilities.

    \section*{Acknowledgements}

    This research received support through Schmidt Sciences, LLC, the International Atomic Energy Authority (IAEA) AI for Fusion Coordinated Research Project and the Imperial College Research Fellowship program.

    This work made extensive use of the Imperial College London RCS HPC systems and the ARCHER2\cite{beckett_2024_14507040} UK National Supercomputing Service (https://www.archer2.ac.uk).

    The authors have benefited from many useful discussions with individuals at I-X, the Australian National University and Data61. We also thank Dr Brian D Appelbe for his helpful review of this work.

    \appendix
    \section{Ensemble Model Uncertainty Calibration}\label{appendix:calibration}

    While an ensemble of ML models can give a distribution of predictions at a point, it is not guaranteed that the statistics of these predictions is representative of the data it was trained/tested on. Often the minimisation of a mean squared error loss ensures the mean prediction is close to the data, however the ensemble variance might be miscalibrated. 
    
    Following Ledda \textit{et al.}\cite{Ledda2023}, we aimed to test our ensemble model calibration by computing the frequency at which test data lies within a given confidence interval of the ensemble model. This is defined mathematically as:
    \begin{align}
        f_\alpha &= \frac{1}{N_D}\sum_{i \in D} I[Y_i > \mu(X_i) - z_\alpha \sigma(X_i) \ \& \   \\
         &Y_i < \mu(X_i) + z_\alpha \sigma(X_i)] \ , \nonumber \\
        z_\alpha &= \Phi^{-1} \left(\frac{1+\alpha}{2}\right)
    \end{align}
    Where $f_\alpha$ is the frequency of test data $D$ in confidence interval $\alpha$, $X_i, Y_i$ are data from test data $D$ which is size $N_D$, $\mu(X)$ and $\sigma(X)$ are the ensemble model predictions of mean and standard deviation at input $X$, $I$ is the indicator function, and $\Phi^{-1}$ is the inverse cumulative probability function of the normal distribution.
    
\begin{figure}[htp]
    \centering
    \includegraphics*[width=0.75\columnwidth]{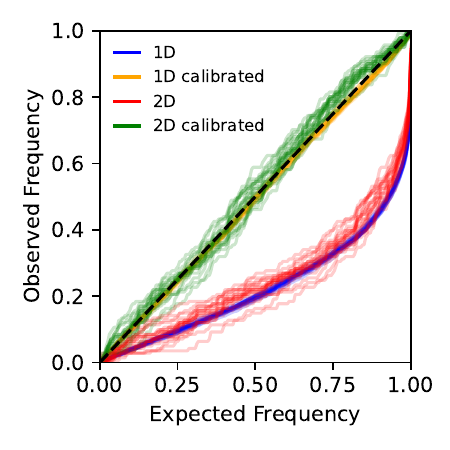}
    \caption{Plot of expected frequency vs observed frequency, $f_\alpha$, for both 1D and 2D ensemble surrogate models, as described in \cref{section:NN}. For each case, 10 random re-samplings (75/25 split) of the complete dataset as used to create testing data. The calibrated curves are created by applying a constant multiplier to $\sigma(X)$ when computing $f_\alpha$.}
    \label{fig:EnsembleCalibration}
    \end{figure}

    Performing this analysis with both our 1D and 2D surrogate ensembles revealed that the trained models were typically `over-confident' in their predictions, i.e. the error bars of the ensemble was too small. \cref{fig:EnsembleCalibration} shows this as the uncalibrated curves lie below the perfect calibration line.
    It was found that applying a constant scale factor ($\sim$ 3.5) on the model uncertainties regained close to perfect calibration. This calibration constant ($c_{v}$) was then included into our ensemble model predictions as follows:
    \begin{align}
        Y_i(X) &\rightarrow \sqrt{c_{v}} Y_i(X) + (1-\sqrt{c_{v}}) \bar{Y}(X)
    \end{align}
    Where $Y_i(X)$ is the $i$-th models prediction at input $X$ and $\bar{Y}(X)$ is the mean prediction over the ensemble. This transformation ensures that the model mean is unaffected but the variances are scaled by $c_{v}$.

    \newpage

    \section*{References}
    \bibliography{MuCFRefs}
	
\end{document}